# Adsorption/desorption and electrically controlled flipping of ammonia molecules on graphene

Shanshan Chen[1,2], Weiwei Cai[1,2]*, David Chen[2], Yujie Ren[2,3], Xuesong Li[2], Yanwu Zhu[2] and Rodney S. Ruoff[2]*

[1]Department of Physics, Fujian Key Laboratory of Semiconductor Materials and Applications, Xiamen University, Xiamen 361005, China

[2]Department of Mechanical Engineering and the Texas Materials Institute, The University of Texas at Austin, Austin, TX 78712 USA

[3]School of Mechanical Engineering, University of Science and Technology Beijing, Beijing, 100083, China

To whom correspondence should be addressed. E-mail:

r.ruoff@mail.utexas.edu (R. S. R.); wc4943@mail.utexas.edu (W. C.)

# Adsorption/desorption and flipping of ammonia molecules on graphene

Abstract:

In this paper, we evaluate of the adsorption/ desorption of ammonia molecules on a graphene surface by studying the Fermi level shift. Based on a physically plausible model, the adsorption and desorption rates of ammonia molecules on graphene have been extracted from the measured Fermi level shift as a function of exposure time. An electric field-induced flipping behavior of ammonia molecules on graphene is suggested, based on field effect transistor (FET) measurements.

## 1. Introduction

Electrical transport experiments on graphene have demonstrated carrier-density-dependent conductivity,[1] the quantum Hall effect,[2] minimum quantum conductivity,[3] and high carrier mobility.[4] Because of these characteristics, graphene is considered a promising new material for memory, logic, analog, opto-electronic, sensor devices, and potentially many other applications.[5-11]

Controlling the intrinsic electrical property and being able to locally change the carrier density are important for graphene devices. It has been shown that graphene is sensitive to molecular adsorbates (e.g. $NH_3$, $H_2O$, $NO_2$ and CO).[12] The Dirac cone band structure of graphene allows control of both the carrier type and the carrier concentration induced by adsorbates due to charge transfer from the adsorbed molecules to graphene. A graphene Hall effect device was capable of sensing individual molecules of $NO_2$.[12] However, the detail of the strength of the adsorption, and the degree of charge transfer for different adsorbates is still debated.[13, 14] In this paper, we report an experimental study on the adsorption/desorption and likely 'flipping' of ammonia molecules on synthetic, large area graphene[15] by detecting the Fermi level shift of a graphene field effect transistor (FET).

**2.　Experiment**

Large-area graphene films grown by chemical vapor deposition ( CVD ) on Cu foils 25-μm thick (Alfa Aesar, item No. 13382)[15] were used to study the adsorption/desorption of $NH_3$ molecules.　The surface of the graphene-on-Cu was first coated with poly-methyl methacrylate (PMMA). After the Cu substrates were dissolved by $Fe(NO_3)_3$ solutions (1M/L), the PMMA-graphene was lifted from the solution and transferred onto a $SiO_2$/Si substrate ($p^+$ doped, ρ~0.002-0.005 Ω-cm; Addison Engineering).[11] Finally, the PMMA was removed by rinsing in acetone at room temperature. Graphene FET devices were constructed by the physical vapor deposition of Au films (~500 nm) as source and drain electrodes on two sides of the graphene film. Figure 1a shows a schematic diagram of the graphene FET used for the transport measurement. Typically, the transport channels defined by the two electrodes deposited on the graphene films were 5-mm wide and 1-mm long.

The quality and the number of stacking layers of the graphene films were determined by micro-Raman spectroscopy (WITec Alpha300, 532 nm laser). Figure 1b shows an optical image (taken at the center of the graphene FET) of the graphene on a $SiO_2$/Si wafer. The 300 nm $SiO_2$/Si wafers are nearly ideal substrates for optically imaging graphene.[16] The uniformity of the color contrast in the optical image indicates uniform graphene thickness, although some small cracks were observed that were likely formed during the transfer process. The Raman spectrum (Figure 1c) shows the following features typical of monolayer graphene: (i) a G-to-2D intensity ratio of ~0.5 and (ii) a symmetric 2D band centered at ~2680 $cm^{-1}$ with a full width at half maximum of ~33 cm$^{-1}$. [15, 17] The D band scattering from our sample, if present at all, was lower than the detection limit of the Raman system used.

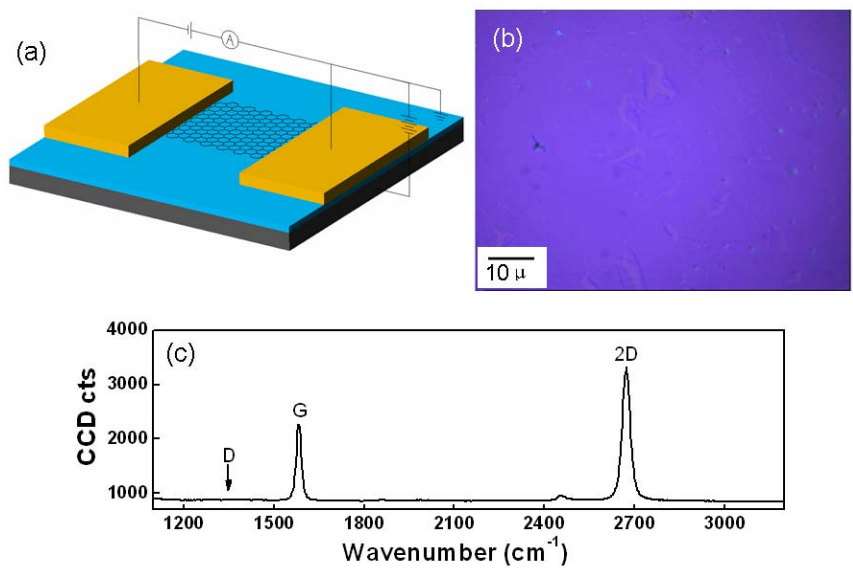

Figure 1 (a) Schematic of a large-area graphene FET supported on a 300-nm thick SiO$_2$-on-Si substrate. Two Au films were deposited as electrodes for the source and drain. The size of the graphene in (a) is about 5×5 mm$^2$. The typical distance between source and drain is 1 mm. Optical image (b) and Raman spectrum (c) of the large-area graphene taken from the center of (b).

Graphene FET devices on SiO$_2$/Si substrates were clamped on a ceramic heater, which was then placed in a high vacuum chamber equipped with electrical and gas feed-throughs. The vacuum chamber can be evacuated to a pressure of 5×10$^{-8}$ Torr using a turbo pump (Varian, Turbo-V 81-M). Vacuum-annealing of the samples was performed *in situ* by heating the graphene on SiO$_2$/Si samples up to 150 °C for 2 hours at 5×10$^{-8}$ Torr in order to eliminate pre-existing adsorbates (i.e., the H$_2$O, O$_2$ molecules). After this vacuum-annealing, the samples were then exposed to NH$_3$ gas (99.99%; Air gas) for known exposure times. FET measurements were performed by a programmable voltage source (Keithley, 2611A) and digital voltmeter/ammeter (Keithley, 6221 and 6514). A back gate bias ($V_{gs}$) ranging from

-100 V to +100 V was applied on the Si side of the $SiO_2$/Si substrate. In order to improve the signal-noise ratio, a relatively high source-drain voltage ($V_{ds}$) of 1.5V was applied to the device while the source-drain current ($I_{ds}$) was monitored as a function of the applied back gate bias.

3. **Results and discussion**

All graphene FETs were measured at room temperature. Figure 2 shows the typical response of the $I_{ds}$ to the gate bias for the as-prepared graphene FET under ambient conditions. A linear $I_{ds}$-$V_{gs}$ curve is observed across the gate bias range used for the as-prepared graphene FET samples. With the gate bias ramped from -80 to 80 V, the $I_{ds}$ decreased from 2.0 to 0.8 mA, indicating that the as-prepared graphene was heavily p-typed; this could be caused by adsorption of water molecules from air, or PMMA residue from the transfer process,[18] or both. Thus, the Dirac point was out of the range of the gate biases that were studied. To minimize the presence of other adsorbates prior to exposure to $NH_3$, the samples were heated at 150 $^o$C for 2 hours under vacuum at $5\times10^{-8}$ Torr. After annealing, a "V" shaped gate response of the $I_{ds}$ is observed from the graphene FET as shown in Figure 2. The *in situ* annealing thus yielded a Dirac point closer to zero gate bias, demonstrating removal of (at least some) adsorbates and the recovery of the intrinsic bias dependence of graphene.

Adsorption/desorption and flipping of ammonia molecules on graphene

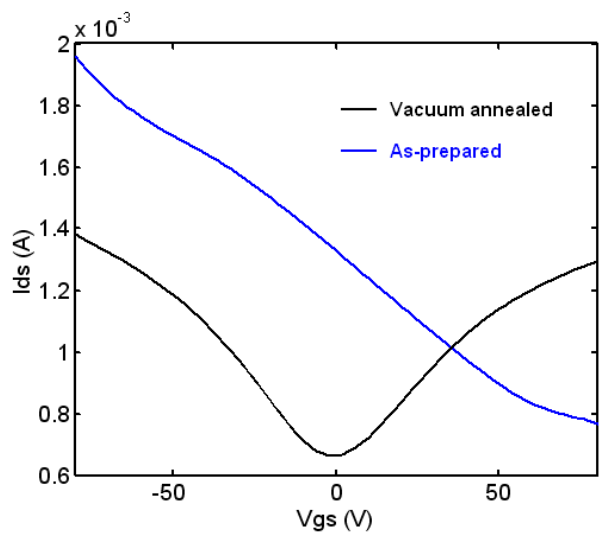

Figure 2 The drain-source current variation of as prepared (blue) and vacuum annealed (black) graphene FET as a function of the Si back gate bias.

In order to study the effect of adsorption of $NH_3$ on the electrical response of graphene, the FET devices were exposed to $NH_3(g)$ after vacuum annealing. The vacuum annealed graphene FET was exposed to 10 Torr of $NH_3$ gas for a total time of 30 minutes. Figure 3a shows the time evolution of the $I_{ds}$ versus $V_{gs}$ during this exposure to $NH_3$ gas. Initially, the Dirac point is close to +3 V back gate bias; after 5 minutes of exposure, the Dirac point appears at -18 V and then gradually shifts to its final position at about -30 V. These results suggest that ammonia molecules adsorb on the graphene surface and cause a shift of the Fermi level in the graphene from the Dirac point into the conduction band.

Research on individual semiconducting carbon nanotubes (CNTs) sensors have been studied based on resistivity changes attributed to molecular adsorption on CNTs and partial electron transfer to the CNTs.[19] A recent first-principles calculation on graphene predicts that dipolar molecules can act as donors with a small charge transfer $\chi$ (e.g. 0.027e for $NH_3$).[13] This is consistent with calculations on defect-free CNTs.[19] Based on this calculated charge transfer of electrons from ammonia to graphene, the number density of the

Adsorption/desorption and flipping of ammonia molecules on graphene

ammonia molecules, $n$, on the graphene surface can be estimated as

$$n\chi = \frac{CV_D}{eS} = \frac{\varepsilon_0 \varepsilon_r V_D}{ed} \quad (1)$$

where $C$ is the capacitance of the graphene FET, $V_D$ is the back gate voltage shift of the Dirac point relative to that of vacuum annealed graphene, $S$ is the surface area, $e$ is the electron charge, $\varepsilon_r$ is the dielectric constant of $SiO_2$, and $d$ is the thickness of $SiO_2$. The values of $n$ as a function of exposure time have been plotted in Figure 3b. The inset in Figure 3b shows the saturated molecular density, with this assumed charge transfer, as a function of the $NH_3$ gas pressure.

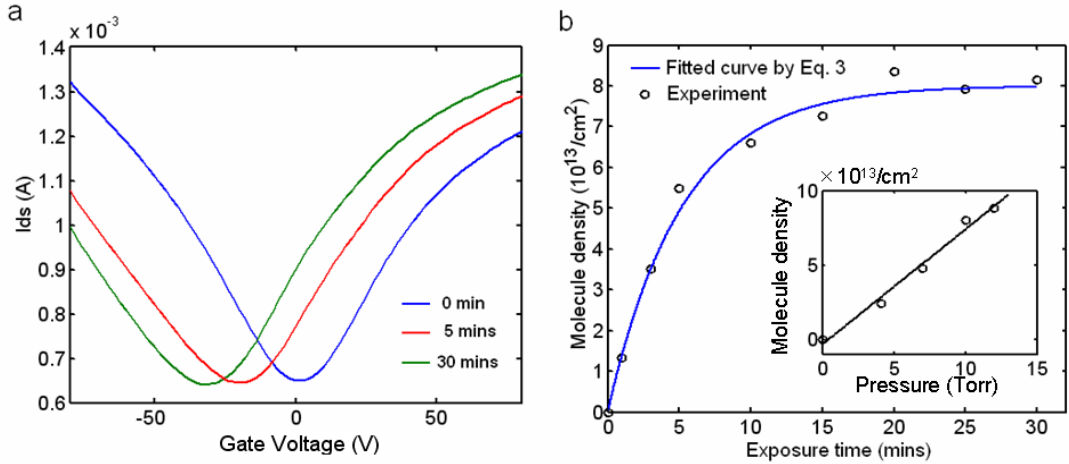

Figure 3 (a) The drain-source current variation of as prepared graphene FET immediately after exposure to $NH_3$ gas for 0, 5, 30 min, respectively. (b) Time evolution of the $NH_3$ molecule density on graphene surface obtained from measurement of $V_D$. The inset in (b) shows the saturated molecule density as a function of $NH_3$ gas pressure.

The adsorption/desorption of the molecules can be understood based on a 'physically plausible' model. Denoting the number density of the molecules in the gas phase as $n_0$ (a constant in our experiments) and the number density on the graphene surface as $n(t)$, the dynamics of the adsorption can be expressed as a rate equation:

Adsorption/desorption and flipping of ammonia molecules on graphene

$$\frac{dn(t)}{dt} \equiv \frac{dn(t)_{ad}}{dt} - \frac{dn(t)_{de}}{dt} = n_0 p_0 - np \quad (2)$$

where $t$ is the exposure time of graphene to NH$_3$ gas, $p_0$ is the adsorption rate and $p$ is the desorption rate. In this model, the number of molecules that adsorb to ($\frac{dn(t)_{ad}}{dt}$) and desorb from ($\frac{dn(t)_{de}}{dt}$) the surface per unit time is assumed to be proportional to the concentration of gas phase ($n_0$), and surface ($n$), molecules, respectively. The adsorption/desorption rates are assumed to be dependent on the temperature only. We also assume that the density of molecules on the surface $n(t) = 0$ at $t = 0$, meaning that the vacuum annealing drives off all adsorbates. By solving Eq. 2, the density of the molecules on the surface as a function of exposure time is obtained as

$$n(t) = \frac{n_0 p_0}{p}(1 - e^{-pt}) \quad (3)$$

A curve fitted to the $n(t)$ data using Eq. 3 is shown in Figure 3 as well. The extracted values of $p$ and $n_0 p_0$ are $0.0027$ and $2.1 \times 10^{10}$ cm$^{-2}$ respectively, and so with the assumptions mentioned above, at a NH$_3$ gas pressure of 10 Torr, about $2.1 \times 10^{10}$ NH$_3$ molecules are adsorbed on 1 cm$^2$ of graphene and 0.27% molecules desorb in one second. It is worth noting that the values of $p$ suggest a sensing response time ($t_0 = 1/p$) if the graphene FET is considered as an NH$_3$(g) sensor.

Many studies, as well as theoretical works,[13, 20] reported that graphene has to be functionalized to achieve its impressive gas-sensing performance. Ab initio studies of gas adsorption onto graphene corroborate the role of impurities or vacancies, thus demonstrating stronger gas adsorption at sites of atomic substitutions or defects.[20] The high sensitivity obtained on reduced graphene oxide gas sensors also supports the importance of functionalization.[21] Recently, experiments compared the electrical gas-sensing performance of "dirty" and intrinsic graphene devices.[14] It was reported that the responses of the intrinsic graphene devices are surprisingly small, even upon exposure to strong analytes such as ammonia vapor. The unintentionally "functionalized" by the residual polymer layer from the lithographic resist served to help concentrate the gas molecules or possibly enhance charge transfer. In our studies, a certain degree of Fermi level shift due to ammonia molecules was detected by performed a FET measurement at room temperature. These results could be caused by the PMMA residue from the transfer process[11] and defects in the graphene grain boundary. Our previous works demonstrate that the typical grain size of CVD grown graphene is ~ 10 μm, which is two orders smaller than the length of the graphene devices.[22]

Another main point of interest is that the extent of charge transfer rate $\chi$ between dipolar molecular adsorbates and graphene could strongly depend on molecular orientation with respect to the graphene surface.[13] The $NH_3$ molecule could, among other possibilities, orient with the N end of the molecule closest to the surface and the $C_{3v}$ axis essentially perpendicular to the surface ("**u**" for "up") or alternatively with the H atoms adjacent to the surface and $C_{3v}$ axis again perpendicular to the surface ("**d**" for "down"). In order to detect if there could be an effect from molecular orientation of the adsorbed $NH_3$ molecules, a scan at low gate bias from -20 V to +20 V was carried out immediately after a high gate bias pulse (+100 V or -100 V) for 5 seconds was applied on the graphene FET devices. Figure 4 shows four sequential $I_{ds}$-$V_{gs}$ measurements in each with the application of one pulse before the scan. The four pulses were applied in the sequence of -100V, +100V, -100V, +100V, respectively. Due to the interaction between the electric field and the molecular dipoles, we suggest that the high positive gate bias pulse aligns $NH_3$ molecules along the **d** orientation, and that the high

Adsorption/desorption and flipping of ammonia molecules on graphene negative bias pulse flips NH$_3$ molecules to the **u** orientation. The low gate bias scan is assumed to have a lesser effect on the orientation of the molecules. The measured curves clearly show that back gate voltages of Dirac point are (repeatably) -6 V (for -100 V pulses) and +3 V (for +100 V pulses). Compared with the vacuum annealed result, $V_D$ for **u** orientation is about -9 V while negligible shift occurs for **d** orientation. This result indicates that, compared with the **d** orientation, the **u** orientation has a relatively large charge transfer ratio, which is consistent with a prediction based on the asymmetry of the highest occupied molecular orbital (HOMO) and lowest unoccupied molecular orbital (LUMO) of the ammonia molecule.[13] The **u** orientation is energetically favored[13] which would explain the donor character observed at zero bias in our experiment. The flipping of some molecular dipoles on graphene could be the cause of the hysteretic behavior reported in other electric field effect measurements.[18]

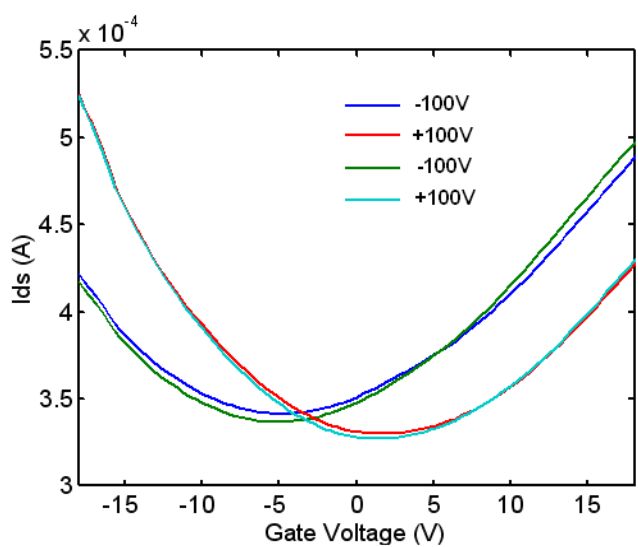

Figure 4 The FET measurement of graphene with ammonia molecules in a low back gate bias range of ±20V following a high back gate bias pulse. The sequence of 5-second pulses: -100 V, +100 V, -100 V, +100 V. Four curves are obtained after each pulse, respectively.

## 4. Summary and conclusions

In this paper, we reported the study of the adsorption/desorption and flipping behavior of ammonia molecules on a graphene surface by observation of the shift in the Fermi level inferred to be from partial charge transfer from the $NH_3$ molecules to graphene. A simple model has been used to evaluate the rates of adsorption and desorption of $NH_3$ molecules on graphene from the measured shift in the Fermi level as a function of exposure time. An electric field induced flip of the molecular dipoles (i.e., the $NH_3$ molecules) is suggested from measured back gate voltage shifts in the Dirac point after electric field pulses were applied via the gate bias.

**Acknowledgement**

Supported by The University of Texas at Austin and by the Texas Nanotechnology Research Superiority Initiative (TNRSI)/SWAN, and support to Shanshan Chen and Yujie Ren from the China Scholarship Council as well as the Special Funds for Major State Basic Research Project and the National Natural Science Foundation of China, are appreciated.

Adsorption/desorption and flipping of ammonia molecules on graphene